\documentclass[twocolumn,amsmath,amssymb,prd,superscriptaddress]{revtex4}
\usepackage{graphicx}

\def\al{\alpha}
\def\be{\beta}
\def\ga{\gamma}
\def\de{\delta}
\def\ep{\epsilon}

\def\et{\eta}
\def\th{\theta}

\def\la{\lambda}

\def\si{\sigma}

\def\ta{\tau}

\def\ph{\phi}

\def\om{\omega}

\def\De{\Delta}

\def\Om{\Omega}

\def\ab{{\al\be}}
\def\prt{\partial}
\def\etal {{\it et al.}}
\newcommand{\beq}{\begin{equation}}
\newcommand{\eeq}{\end{equation}}
\newcommand{\bea}{\begin{eqnarray}}
\newcommand{\eea}{\end{eqnarray}}
\newcommand{\bse}{\begin{subequations}}
\newcommand{\ese}{\end{subequations}}

\def\ni{\noindent}
\def\sF#1#2{{\textstyle{{#1}\over{#2}}\,}}
\newcommand{\BB}{\big}
\newcommand{\nn}{\nonumber}

\def\ol#1{\overline{#1}}

\usepackage{amsmath}
\usepackage{amsfonts}
\usepackage{amssymb}

\newcommand{\bM}{\begin{pmatrix}}
\newcommand{\eM}{\end{pmatrix}}

\newcommand{\WT}{\Om_\odot T}
\newcommand{\pp}{\mathbf{p}}
\newcommand{\MM}{\text{M}}
\def\Re{\hbox{Re}}

\newcommand{\ha}{\sF{1}{2}}

\def\vev#1{\langle {#1}\rangle}

\def\T#1#2{\ta^{(#1)}_{#2}}
\def\S#1#2{S^{(#1)}_{#2}}
\def\gt{\widetilde g} 
\def\Ht{\widetilde H}

\def\nub{\bar{\nu}}

\usepackage{slashed}
\def\sl#1{\slashed{#1}}

\begin{document}

\title{Limits on CPT violation from solar neutrinos}
\author{Jorge S. D\'iaz}
\affiliation{Institute for Theoretical Physics, Karlsruhe Institute of Technology (KIT), 76128 Karlsruhe, Germany}
\author{Thomas Schwetz}
\affiliation{Institute for Nuclear Physics, Karlsruhe Institute of Technology (KIT), 76021 Karlsruhe, Germany}

\begin{abstract}
Violations of CPT invariance can induce neutrino-to-antineutrino transitions. 
We study this effect for solar neutrinos and use the upper bound on the solar neutrino-to-antineutrino transition probability from the KamLAND experiment to constrain CPT-symmetry-violating coefficients of the general Standard-Model Extension. 
The long propagation distance from the Sun to the Earth allows us to improve existing limits by factors ranging from about a thousand to $10^{11}$.
\end{abstract}

\maketitle


\section{Introduction}

After escaping the Sun, neutrinos propagate in vacuum approximately 150 million km before reaching our detectors on Earth \cite{SolarExps01,SolarExps02,SolarExps03,SolarExps04,SolarExps05,SolarExps06,SolarExps07}. 
Although solar neutrinos of all three flavors are expected to reach us due to flavor mixing, 
no antiparticle counterparts are expected. 
Borexino \cite{Borexino:2011}, KamLAND \cite{KamLAND:2004,KamLAND:2012}, SNO \cite{SNO:2004}, and Super-Kamiokande \cite{SK:2003} have performed systematic searches for electron antineutrinos ($\nub_e$) coming from the Sun.
The absence of a positive signal has been used to constrain the parameters of models beyond standard mass-flavor mixing, which could trigger the transition of a neutrino into an antineutrino.

Deviations from exact Lorentz invariance could produce several unconventional effects in neutrino experiments \cite{Diaz:2014c}.
One interesting effect is the mixing between neutrinos and antineutrinos triggered by Majorana couplings in the Standard-Model Extension (SME) \cite{SME1,SME2,SME3}.
This mixing can lead to different experimental signatures.
Limits on some of the controlling coefficients have been obtained from neutrinoless double beta decay experiments \cite{Diaz:2014b},
and neutrino-antineutrino oscillations have been explored in accelerator and reactor experiments \cite{Rebel:2013,Diaz:2013b}.

In this work,
we use the results from solar-neutrino experiments to determine the most stringent limits on the coefficients for Lorentz and CPT violation that would produce $\nu_e\to\nub_e$ oscillations.
The long propagation distance makes solar neutrinos highly sensitive to minuscule effects that are enhanced by the baseline.

This paper is organized as follows. In Sec. \ref{Sec.Theory} the general theory of Lorentz-violating neutrinos is reviewed, 
while the relevant transition probability is determined in Sec. \ref{Sec.Probability}.
In Sec. \ref{Sec.RefFrame} the choice of reference frame is discussed so the theory can be applied in Sec. \ref{Sec.Application}.
Section \ref{Sec.Summary} concludes with a summary.

\section{Theory}
\label{Sec.Theory}

Lorentz-violating neutrinos and antineutrinos in the SME are effectively described by the Lagrangian density \cite{KM:2004,KM:2012}

\beq\label{L_nu}
\mathcal{L} = \ha \ol\Psi\BB(i \sl{\prt}- M + \hat{\mathcal{Q}}\BB)\Psi + \text{h.c.},
\eeq
where the multiplet $\Psi=(\nu_e,\nu_\mu,\nu_\tau,\nu_{e}^C,\nu_{\mu}^C,\nu_{\tau}^C)^T$ includes the states of three neutrinos and their charge conjugates,
$M$ is a mass matrix,
and the generic operator for Lorentz violation $\hat{\mathcal{Q}}$ can be decomposed in a basis of Dirac matrices in the general form \cite{KM:2012}
\bea
\hat{\mathcal{Q}} &=& \hat{\mathcal{S}} + i\hat{\mathcal{P}} \ga_5 + \hat{\mathcal{V}}^\la \ga_\la 
    + \hat{\mathcal{A}}^\la \ga_5\ga_\la  + \ha \hat{\mathcal{T}}^{\la\et} \si_{\la\et}.
\eea
Each of the elements in this expansion is a $6\times6$ matrix,
which can be further decomposed into $3\times3$ blocks of Dirac and Majorana type \cite{KM:2012}.
Dirac components affect neutrinos and antineutrinos independently,
whereas Majorana components induce neutrino-antineutrino mixing.
From the Lagrangian \eqref{L_nu} an effective Hamiltonian can be constructed. 
A seesaw mechanism for neutrino masses is then implemented, 
which suppresses the propagation of sterile states.
The full Hamiltonian incorporates a conventional Lorentz-invariant part and
the Lorentz-violating piece.
The latter involves only the Dirac part of the vector $\hat{\mathcal{V}}^{\la}$ and axial vector $\hat{\mathcal{A}}^{\la}$ components as well as the Majorana part of the tensor component $\hat{\mathcal{T}}^{\la\et}$;
scalar $\hat{\mathcal{S}}$ and pseudoscalar $\hat{\mathcal{P}}$ components are irrelevant at leading order \cite{KM:2012}.
The Dirac-type terms modify the propagation and mixing of neutrinos and antineutrinos independently in the form of two coefficients denoted $(a_L)^\la$ and $(c_L)^{\la\et}$,
which control CPT-odd and CPT-even Lorentz violation, respectively.
These coefficients have been studied in a variety of accelerator, atmospheric, and reactor oscillation experiments \cite{LV_LSND, LV_MINOS_ND1, LV_IceCube, LV_MINOS_FD, LV_MiniBooNE:2012, LV_DoubleChooz, LV_MINOS_ND2, LV_MiniBooNE:2013,LV_SuperK}
and a recent study searched for key signatures in double beta decay \cite{LV_EXO:2016}.
All experimental results are tabulated in Ref. \cite{DataTables}.

The relevant tensor component of the effective Hamiltonian depends on two coefficients denoted $\gt^{\la\et}$ and $\Ht^\la$,
which control CPT-odd and CPT-even Lorentz violation, respectively.
These coefficients produce the mixing of left-handed neutrinos and right-handed antineutrinos,
leading to the possibility of neutrino-antineutrino oscillations.
This experimental signature has been studied using accelerator neutrinos in the MINOS experiment \cite{Rebel:2013} and antineutrinos in the Double Chooz reactor experiment \cite{Diaz:2013b}.

We remark in passing that in realistic field theories, 
operators that break CPT invariance are a subset of those that break Lorentz symmetry \cite{Greenberg:2002}.
For this reason,
hereafter when we refer to CPT violation, 
it is understood that Lorentz invariance is also broken.

The relevant Hamiltonian is given by \cite{KM:2004,Diaz:2009}
\beq\label{H1a}
\de H_{\bar\al\be} = i\sqrt2 (\ep_+)^*_\la \BB[\,\hat p_\si E \, \gt^{\la\si}_{\bar\al\be} - \Ht^\la_{\bar\al\be} \,\BB],
\eeq
where the generation indices $\bar{\al}\in\{\bar{e},\bar{\mu},\bar{\tau}\}$, 
$\be\in\{e,\mu,\tau\}$ have been included in the coefficients for Lorentz violation $\gt^{\la\si}_{\bar\al\be}$ and $\Ht^\la_{\bar\al\be}$.
The other quantities in this Hamiltonian are the neutrino energy $E$,
$(\ep_+)_\la$ is a polarization 4-vector,
and $\hat p_\si=(1,-\hat\pp)$ is a parametrization of the neutrino direction of propagation.
Due to the Hermiticity of the full $6\times6$ Hamiltonian and its transformation properties under charge conjugation,
the coefficients for CPT-even Lorentz violation are antisymmetric in the mixed flavor space 
$\Ht^\la_{\bar\al\be} =-\Ht^\la_{\bar\be\al}$.
Similarly, 
the dimensionless coefficients for CPT-odd Lorentz violation are symmetric in the mixed flavor space 
$\gt^{\la\si}_{\bar\al\be} = \gt^{\la\si}_{\bar\be\al}$ \cite{KM:2004}.
These symmetry properties will play an important role in the evaluation of the oscillation probability in Sec. \ref{Sec.Probability}.
In fact,
it will be shown that the oscillation probability involves a symmetric combination of coefficients,
making only the symmetric part of the Hamiltonian relevant.
In other words,
the coefficients $\Ht^\al_{\bar\al\be}$ are unobservable in the oscillation channel $\nu_j\to\nub_e$.
For this reason,
we will rewrite the Lorentz-violating Hamiltonian \eqref{H1a} simply as
\beq\label{H1}
\de H_{\bar\al\be} = i\sqrt2 (\ep_+)^*_\la \, \hat p_\si E \, \gt^{\la\si}_{\bar\al\be}.
\eeq

\section{Oscillation probability}
\label{Sec.Probability}

The appearance of electron antineutrinos from the Sun could be interpreted as caused by Lorentz violation,
which could make a left-handed electron neutrino oscillate into a right-handed antineutrino.
The absence of a signal of the transition $\nu_e\to\nub_e$ over such a long propagation distance implies that the relevant coefficients for Lorentz violation are very small.

The description of neutrinos moving through the Sun requires the incorporation of matter effects.
For the first $\sim700\,000$ km,
the propagation through a medium of electron density $n_e(r)$ is dominated by the matter potential $V(r)=\sqrt2\,G_Fn_e(r)$ that modifies the mixing angles,
making them a function of the vacuum mixing angles, the neutrino energy, and position through the matter potential.
Given the existing constraints on the coefficients $\gt^{\la\si}_{\bar\al\be}$ \cite{Rebel:2013,Diaz:2013b},
CPT-violating effects in the propagation through the solar interior can be safely neglected.
The role of $\gt^{\la\si}_{\bar\al\be}$ is to produce unconventional transitions in the vacuum propagation of neutrinos to Earth.
Therefore,
the $\nu_e\to\nub_e$ transition probability can be factorized into two terms,
one describing the neutrino propagation from the solar core to the surface,
and the other characterizing the evolution in vacuum from the surface of the Sun to Earth in the form
\beq\label{Peebar}
P_{\nu_e\to\nub_e} = \sum_j P^\odot_{\nu_e\to\nu_j} \, P^\text{LV}_{\nu_j\to\nub_e},
\eeq
where we have used that the sum over the three mass eigenstates is incoherent due to averaging of fast oscillations. 
Each of the two terms in \eqref{Peebar} is determined in the next subsections.

\subsection{From $r=0$ to $r=R_\odot$}

Inside the Sun,
the adiabatic evolution of the mass eigenstate in matter $\nu_i^\MM$ guarantees that at $r=0$ this state coincides with the mass eigenstate evolved to the solar surface $r=R_\odot$, where the vacuum mixing is recovered $\nu_i^\MM=\nu_i$.
This leads to the standard conversion probability,
\beq\label{P_ej(Sun)}
P^\odot_{\nu_e\to\nu_j} = |U_{ej}^\MM|^2 ,
\eeq
where the mixing matrix is evaluated at $r=0$.
This is an approximation;
however,
an average within the production region leaves our results unchanged so it is neglected.
Since $V(0)\ll \De m^2_{31}/E$,
the mixing angle $\th_{13}$ is not enhanced by matter effects and we can take $\th_{13}^\MM\approx\th_{13}$ in the expression above to write the relevant components of the mixing matrix as
\bea
|U_{e1}^\MM|^2 &=& \cos^2\th_{12}^\MM\,\cos^2\th_{13}^\MM
\approx \frac{1}{2}\cos^2\th_{13} \BB(1+\cos2\th_{12}^\MM\BB),
\nn\\
|U_{e2}^\MM|^2 &=& \sin^2\th_{12}^\MM\,\cos^2\th_{13}^\MM
\approx \frac{1}{2}\cos^2\th_{13} \BB(1-\cos2\th_{12}^\MM\BB),
\nn\\
|U_{e3}^\MM|^2 &=& \sin^2\th_{13}^\MM \approx \sin^2\th_{13},
\eea
where
\beq
\cos2\th_{12}^\MM =
\frac{\cos2\th_{12}-2EV_0/\De m^2_{21}}{\sqrt{\big( \cos2\th_{12}-2EV_0/\De m^2_{21}\big)^2 + \sin^22\th_{12}}} .
\label{U_e1M}
\eeq
The matter potential at the solar core is given by $V_0=7.84\times10^{-21}$ GeV \cite{Bahcall:2006}.

\subsection{From $r=R_\odot$ to $r=L$}

After a neutrino state leaves the Sun,
its evolution is solely controlled by vacuum oscillations.
The full Hamiltonian can be written as the sum of a conventional Lorentz-invariant part and a small modification introduced by Lorentz violation in the form
\beq
H = H_0 + \de H.
\eeq
Each term is a $6\times6$ matrix describing the evolution of three left-handed neutrinos and three right-handed antineutrinos.
The Lorentz-invariant part has a block-diagonal form,
whereas in $\de H$ the Majorana part of the tensor component appears in the off-diagonal $3\times3$ block.

Since large Lorentz violation remains unobserved,
any deviation from exact Lorentz symmetry is expected to be small.
We use this argument to treat Lorentz violation perturbatively.
For the appropriate implementation of perturbation theory,
the unperturbed Hamiltonian must be diagonalized first.
Since the unperturbed Hamiltonian corresponds to conventional vacuum oscillations,
the eigenenergies can be written as
\beq\label{E0_123}
E^0_k = E^0_{\bar k} \approx |\pmb{p}| + \frac{m^2_k}{2|\pmb{p}|},
\eeq
where the index $k\in\{1,2,3\}$ denotes neutrino states and $\bar k\in\{\bar 1,\bar 2,\bar 3\}$ denotes antineutrino states.
The CPT theorem implies that the Lorentz-invariant eigenenergies of $H_0$ are the same for neutrinos and antineutrinos \cite{Greenberg:2002}.
The mixing matrix $\mathcal{U}$ for the unperturbed system contains the conventional PMNS matrix $U$ as diagonal blocks
\beq\label{U_matrix}
\mathcal{U} = \bM U & 0 \\ 0 & U^* \eM.
\eeq

A perturbative series can now be implemented following the procedure in Ref. \cite{Diaz:2009}.
The time-evolution operator takes the form
\bea
S(t) &=& e^{-iHt} \nn\\
&=& e^{-iHt}\,e^{iH_0t} \S{0}{}(t)
= W(t) \, \S{0}{}(t)\nn\\
&=& \S{0}{}(t) + \S{1}{}(t) + \S{2}{}(t) + \cdots,
\label{S(series)}
\eea
where $\S{n}{}(t)$ is the term to order $n$th in the perturbation $\de H$ and we have used
$\S{0}{}(t)=e^{-iH_0t}$.
The function $W(t)$ can be expressed as a Dyson series
\bea
W(t) &=& W^{(0)} + W^{(1)} + W^{(2)} + \cdots \nn\\
 &=&  1 +(-i)\int_0^t\!\!dt_1 \BB(e^{-iH_0t_1}\de H e^{iH_0t_1} \BB) \nn\\
 &&
      +(-i)^2\int_0^t\!\!dt_2 \int_0^{t_2}\!\!dt_1 \BB(e^{-iH_0t_1}\de H e^{iH_0t_1}\BB) \nn\\
 && \qquad \times\BB(e^{-iH_0t_2}\de H e^{iH_0t_2}\BB)
      +\cdots.
\eea
Since in the mass basis the unperturbed Hamiltonian $H_0$ is diagonal,
the first two terms in the series for $W(t)$ in this basis become
\bea
W^{(0)}_{JK} &=& \de_{JK},
\\
W^{(1)}_{JK} &=& -i\int_0^t\!\!dt_1 \BB(e^{-iE^{(0)}_J\de_{JL}t_1}\de H_{LM} e^{iE^{(0)}_M\de_{MK}t_1} \BB)
\nn\\
&=& -i\int_0^t\!\!dt_1 \BB(e^{-i(E^{(0)}_J-E^{(0)}_K)t_1}\de H_{JK} \BB)
\nn\\
&=& -it \,\de H_{JK}\, \tau^{(1)}_{JK}(t)\,e^{iE^{0}_Kt},
\eea
where the uppercase indices $J,K\in\{1,2,3,\bar1,\bar2,\bar3\}$ span the eigenstates of neutrinos and antineutrinos.
In the above expressions,
$E^{0}_J$ are the eigenvalues of the unperturbed Hamiltonian $H_0$ given in \eqref{E0_123} and we have defined
\beq
\tau^{(1)}_{JK}(t) = \left\{\begin{array}{ccc} e^{-iE^{0}_Jt} &,& E^{0}_J = E^{0}_K \\
\dfrac{e^{-iE^{0}_Jt} - e^{-iE^{0}_Kt}}{-i(E^{0}_J-E^{0}_K)t} &,& E^{0}_J \neq E^{0}_K
\end{array}\right..
\eeq
The first two elements of the expansion for the time-evolution operator \eqref{S(series)} become
\bea
\S{0}{JK} &=& \de_{JK}\,e^{-iE^{0}_Jt},
\nn\\
\S{1}{JK} &=& -it \,\de H_{JK}\, \tau^{(1)}_{JK}(t).
\label{S0_S1}
\eea
We are interested in the transition probability from a mass eigenstate $\nu_j$ (with $j=1,2,3$) to the flavor state $\nub_e$,
which is given by
\bea
P^\text{LV}_{\nu_j\to\nub_e} &=& \BB| S_{\bar ej} \BB|^2
= \BB| \S{0}{\bar ej} + \S{1}{\bar ej} + \S{2}{\bar ej} + \cdots \BB|^2
\nn\\
&=& \BB| \S{0}{\bar ej} \BB|^2 + 2\,\Re\BB(S^{(0)*}_{\bar ej}\S{1}{\bar ej} \BB)
\nn\\
&& + 2\,\Re\BB(S^{(0)*}_{\bar ej}\S{2}{\bar ej} \BB) + \BB|\S{1}{\bar ej} \BB|^2 + \cdots ,
\label{P_je}
\eea
where only terms up to second order are explicitly shown.
Using the mixing matrix \eqref{U_matrix} to relate the mass and flavor bases,
the relevant components of the time-evolution operator \eqref{S(series)} can be split into a sum over neutrino and antineutrino states in the form 
\beq\label{S_ej}
S_{\bar ej} = \sum_{K} \mathcal{U}_{\bar eK} \,S_{Kj}
         = \sum_{k} \mathcal{U}_{\bar ek} \,S_{kj} + \sum_{\bar k} \mathcal{U}_{\bar e\bar k} \,S_{\bar kj},
\eeq
where $k\in\{1,2,3\}$ and $\bar k\in\{\bar 1,\bar 2,\bar 3\}$, 
and we have taken $J=j\in\{1,2,3\}$ because only neutrino mass eigenstates are produced in the Sun.
The first term in \eqref{S_ej} vanishes because the $6\times6$ mixing matrix \eqref{U_matrix} is block diagonal $\mathcal{U}_{\bar ek} = 0$.
Applying the transformation \eqref{S_ej} order by order to the terms in \eqref{S0_S1}, we find 
\bea
\S{0}{\bar ej} &=& \sum_{\bar k} \mathcal{U}_{\bar e\bar k} \S{0}{\bar kj}
 = \sum_{\bar k} \mathcal{U}_{\bar e\bar k} \,\de_{j\bar k}\,e^{-iE^{0}_jt} = 0,
\nn\\
\S{1}{\bar ej} &=& \sum_{\bar k} \mathcal{U}_{\bar e\bar k} \S{1}{\bar kj}
 = -it \,\sum_{\bar k} \mathcal{U}_{\bar e\bar k} \,\de H_{\bar kj}\, \tau^{(1)}_{\bar kj}
 \nn\\
&=& -it \,\sum_{\bar\al\be} \sum_{\bar k} \tau^{(1)}_{\bar kj} U_{\bar e\bar k} \,U^*_{\bar\al\bar k} U_{\be j} \,\de H_{\bar\al\be},
\label{S1b}
\eea
where the Lorentz-violating Hamiltonian has been written in the flavor base,
with $\bar\al\in\{\bar e,\bar\mu,\bar\ta\}$, $\be\in\{e,\mu,\ta\}$.
At leading order the probability \eqref{P_je} reduces to
\beq\label{P_jebar(LV)}
P^\text{LV}_{\nu_j\to\nub_e} = \BB| \S{1}{\bar ej} \BB|^2.
\eeq

The long propagation distance compared to the relevant oscillation length allows us to simplify some expressions.
This occurs due to the decoherence of the neutrino wave packet,
making some oscillatory terms average to zero.
The amplitude $\S{1}{\bar ej}$ given by \eqref{S1b} can be expressed in the form
\beq\label{S1c}
\S{1}{\bar ej} = -it \,e^{-iE^0_jt}\,\sum_{\bar\al\be} \sum_{\bar k} e^{iE^0_jt}\,\tau^{(1)}_{\bar kj} U_{\bar e\bar k} \,U^*_{\bar\al\bar k} U_{\be j} \,\de H_{\bar\al\be},
\eeq
where the product of the exponential and the functions $\T{1}{\bar kj}$ for a large propagation distance $L\simeq t$ compared to the oscillation lengths become 
\begin{align}
e^{iE^0_jt}\,\tau^{(1)}_{\bar kj} &= 1
, & E^{0}_j = E^{0}_{\bar k},
\nn\\
e^{iE^0_jt}\,\tau^{(1)}_{\bar kj} 
&\ll 1
, & E^{0}_j \neq E^{0}_{\bar k}.
\end{align}
This result shows that only the diagonal elements of $\tau^{(1)}_{\bar kj}$ contribute.
For solar neutrinos, 
which after leaving the Sun propagate around $L\sim1.5\times10^{8}$ m,
the off-diagonal components are of order $\mathcal{O}(10^{-6})$ or less.
Using the CPT theorem for relating neutrino and antineutrino indices, 
the product of the exponential and the functions $\T{1}{\bar kj}$ becomes a simple Kronecker delta $\de_{jk}$.
Expression \eqref{S1c} finally can be written as
\beq
\S{1}{\bar ej} = -it \,e^{-iE^0_jt}\,U^*_{ej} \sum_{\al\be} \, U_{\al j} U_{\be j} \,\de H_{\bar\al\be},
\eeq
and the transition probability \eqref{P_jebar(LV)} for neutrinos in vacuum takes the form
\beq\label{P_jebar(LV)2}
P^\text{LV}_{\nu_j\to\nub_e} = L^2 
\Big| U_{ej} \sum_{\al\be}  \,U_{\al j} U_{\be j} \,\de H_{\bar\al\be} \Big|^2 .
\eeq
This transition probability together with the result in \eqref{P_ej(Sun)} is what we need to construct the full transition probability for $\nu_e\to\nub_e$ defined in \eqref{Peebar}.
As mentioned at the end of Sec. \ref{Sec.Theory},
the oscillation probability involves a symmetric combination of coefficients,
making the antisymmetric coefficients $\Ht^\al_{\bar\al\be}$ unobservable in the oscillation channel $\nu_j\to\nub_e$.
For the explicit application of the probability \eqref{P_jebar(LV)2}, 
we need to specify the reference frame in which the Hamiltonian $\de H_{\bar\al\be}$ is expressed.
These details are described in the following section.

\section{Reference frame}
\label{Sec.RefFrame}

The appropriate application of the result \eqref{P_jebar(LV)2} requires a choice of reference frame.
Even though Lorentz symmetry is broken,
invariance under coordinate transformations, 
also known as {\it observer Lorentz transformations},
remains unchanged \cite{SME1,SME2,Diaz:2014c}.
This means that any observer frame is equally valid to express the relevant Hamiltonian \eqref{H1a}.
In order to report experimental results in a meaningful way,
which can be used to compare with other experiments,
a convention is required for the frame of reference in which measurements are made.
The standard frame used is the Sun-centered equatorial frame \cite{SunFrame1,SunFrame2,SunFrame3}.
In this frame,
the $\hat X$ axis points towards the vernal equinox from the Sun,
while the axis of rotation of the Earth determines the $\hat Z$ axis.
The $\hat Y$ axis completes the system as $\hat Y=\hat Z\times\hat X$.
For experiments with both source and detector on the surface of the Earth,
the neutrino propagation oscillates with sidereal frequency $\om_\oplus\simeq2\pi/(\text{23 h 56 m})$.
This oscillation is then used to decompose the relevant observable in harmonics of the sidereal phase $\om_\oplus T_\oplus$ \cite{KM:2004b,Diaz:2009}.
This technique has been widely used in searches for Lorentz violation in oscillation experiments \cite{Rebel:2013,LV_LSND, LV_MINOS_ND1, LV_IceCube, LV_MINOS_FD, LV_MiniBooNE:2012, LV_DoubleChooz, LV_MINOS_ND2, LV_MiniBooNE:2013}.

In the case of solar neutrinos,
only the detector is on the Earth,
while the source is fixed at the origin of the coordinate system.
A different approach is then required for an appropriate treatment of the oscillation probability.
The Sun continuously emits neutrinos in all directions;
however,
we are only interested in those neutrinos emitted in the direction $\hat\pp$, 
defined as the source-detector orientation.
As the Earth moves around the Sun, 
the vector $\hat\pp$ changes with respect to the fixed coefficients for Lorentz violation,
hence a time dependence of the oscillation probability will arise.
Instead of sidereal time,
a more reasonable time choice is the use of solar time.
The origin of the time coordinate in the Sun-centered equatorial frame is the vernal equinox in the year 2000 \cite{DataTables}.
For this reason,
time must be measured with respect to this event that defines $T=0$.

In the Sun-centered equatorial frame,
the source-detector orientation is given by
\beq
\hat\pp = \BB(-\cos\WT, -\cos\eta \sin\WT, -\sin\eta \sin\WT \BB),
\eeq
where $\Om_\odot\simeq 2\pi/(\text{365.25 d})$ is the annual frequency of the Earth around the Sun and $\eta\simeq 23.5^\circ$ denotes the inclination of the orbital  plane with respect to the plane of the celestial equator.
Introducing a spherical basis with $\hat e_r=\hat\pp$,
we can write the two vectors, 
\bea
\hat e_\th &=& \BB(\sin\WT, -\cos\eta \cos\WT, -\sin\eta \cos\WT \BB), 
\nn\\
\hat e_\ph &=& \BB(0, \sin\eta , -\cos\eta \BB),
\eea
to form an orthonormal basis.
The helicity vector in \eqref{H1} is given by $(\ep_+)_\la=(0,-\vec\ep_+)$ with
\beq
\vec\ep_+ =\frac{1}{\sqrt{2}}\BB(\hat{e}_\th + i \hat{e}_\ph \BB)  .
\eeq
The relevant 4-vectors in expression \eqref{H1} in the Sun-centered equatorial frame take the explicit form
\bea
(\ep_+)_\la &=& \frac{1}{\sqrt{2}} 
\BB(0, -\sin\WT, \cos\eta \cos\WT - i\sin\eta, \nn\\
&&\qquad \sin\eta \cos\WT + i\cos\eta \BB),
\nn\\
\hat{p}_\si &=& \BB(1, \cos\WT, \cos\eta \sin\WT, \sin\eta \sin\WT \BB).
\qquad
\eea
We can define the time-dependent functions as
\beq
f_{\la\si} = i\sqrt2 (\ep_+)^*_\la \, \hat p_\si,
\eeq
which take a definite form for each spacetime component of the coefficient of interest.
The 12 nonvanishing components are explicitly given by
\bea
f_{XT} &=& -i \sin\WT ,\nn\\
f_{XX} &=& -\sF{i}{2} \sin{2\WT} ,\nn\\
f_{XY} &=& -i \cos\eta\,\sin^2\WT ,\nn\\
f_{XZ} &=& -i \sin\eta\sin^2\WT ,\nn\\
f_{YT} &=&  i \cos\eta\,\cos\WT-\sin\eta ,\nn\\
f_{YX} &=& \cos\WT \BB(i \cos\eta\,\cos\WT-\sin\eta \BB) ,\nn\\
f_{YY} &=& \cos\eta\,\sin\WT\, \BB(i \cos\eta\,\cos\WT-\sin\eta \BB) ,\nn\\
f_{YZ} &=& \sin\eta\,\sin\WT\, \BB(i \cos\eta\,\cos\WT-\sin\eta \BB) ,\nn\\
f_{ZT} &=& i \sin\eta\,\cos\WT + \cos\eta,\nn\\
f_{ZX} &=& \cos\WT\, \BB(i \sin\eta\,\cos\WT + \cos\eta \BB) ,\nn\\
f_{ZY} &=& \cos\eta\,\sin\WT\, \BB(i \sin\eta\,\cos\WT + \cos\eta \BB) ,\nn\\
f_{ZZ} &=& \sin\eta\,\sin\WT\,(i \sin\eta\,\cos\WT + \cos\eta \BB).
\eea
\ni
The oscillation probability \eqref{P_jebar(LV)2} can now be written in terms of the above functions in the form
\beq\label{PeebarLV(E,T)}
P^\text{LV}_{\nu_j\to\nub_e}(E,T) 
= L^2(T) E^2 \, \bigg| U_{ej}\, f_{\la\si}(T) \sum_{\al\be}  \,U_{\al j} U_{\be j} \, \gt^{\la\si}_{\al\bar\be} \bigg|^2,
\eeq
where the time and energy dependence of each term has been explicitly displayed 
and $\gt^{\la\si}_{\bar\al\be}$=$\gt^{\la\si*}_{\al\bar\be}$ has been used \cite{Diaz:2009}.
The propagation distance depends on time due to the nonzero eccentricity of Earth's orbit in the form 
\beq
L(T) = \frac{a(1-e^2)}{1+e\,\cos\th(T)} - R_\odot,
\eeq
where $a=1.50\times10^8$ km is the Earth semimajor axis,
$e=0.02$ is the eccentricity of the orbit \cite{OrbitalParam},
and $R_\odot=6.96\times10^5$ km is the solar radius.
The polar angle $\th(T)$ is measured from the perihelion,
which in the year 2000 occurred on January 3, 2000, 5:18 GMT \cite{NavalObs}.
We can then write the polar angle as a function of time in the form
\beq\label{theta(T)}
\th(T) = \th_0 + \WT,
\eeq
where the constant angle $\th_0$ is the difference between the perihelion and the vernal equinox of the year 2000.
The vernal equinox occurred on March 20, 2000, 7:35 GMT \cite{NavalObs};
hence, 
$\th_0=\Om_\odot\De T$, 
with $\De T=77.095$ d.

\section{Application to KamLAND}
\label{Sec.Application}

We can now apply the previous results to experimental values.
The most sensitive limit on $\nu_e\to\nub_e$ conversion was obtained by KamLAND \cite{KamLAND:2012};
hence, 
we will use this result to determine limits of the coefficients for CPT violation $\gt^{\la\si}_{\bar ab}$.
The experimental limit on the transition probability is \cite{KamLAND:2012}
\beq\label{Pee_exp}
\vev{P_{\nu_e\to\nub_e}}_{\text{exp}} < 5.3 \times 10^{-5}\qquad \text{(90\% C.L.)}.
\eeq
Notice that the oscillation probability has been averaged over the energy range 8.3 MeV $<E<$ 31.8 MeV and the analysis includes data accumulated between March 5, 2002, and July 23, 2010 \cite{KamLAND:2012}.
Since the exact time of day in which the data collection took place is unavailable,
we will assume that the period began and ended at local midnight of the corresponding date.
Our final results, 
however,
have little dependence on this assumption.
With this choice,
we will use the local time $t$ given by a shift on $T$ to compensate for the different time zones:
\beq
T(t) = t - 0.691 \text{ d}.
\eeq
We can now properly average the oscillation probability \eqref{Peebar} over time and energy in the form
\beq\label{<Peebar>}
\vev{P_{\nu_e\to\nub_e}} = 
\frac{\displaystyle\int_{t_1}^{t_2} \!\! dt \displaystyle\int_{E_1}^{E_2}\!\! dE \,\si(E)\ph(E)P_{\nu_e\to\nub_e}(E,t)}
{\displaystyle\int_{t_1}^{t_2} \!\! dt \displaystyle\int_{E_1}^{E_2}dE \,\si(E)\ph(E)},
\eeq
with $t_1=715$ d, $t_2=3777$ d, $E_1=8.3$ MeV, and $E_2=31.8$ MeV.
The energy average includes the inverse beta decay cross section $\si(E)$ \cite{Vogel:1999} and the energy spectrum of $^8$B neutrinos $\ph(E)$ \cite{Bahcall:1996}.
Following the approach implemented in previous experimental searches \cite{Rebel:2013,Diaz:2013b,LV_LSND, LV_MINOS_ND1, LV_IceCube, LV_MINOS_FD, LV_MiniBooNE:2012, LV_DoubleChooz, LV_MINOS_ND2, LV_MiniBooNE:2013,LV_SuperK},
we will consider that each individual coefficient $\gt^{\la\si}_{\al\bar\be}$ is independently small.
This means that for a given set of spacetime components $\la\si$ and flavor indices $\al\bar\be$,
we will consider each coefficient $\gt^{\la\si}_{\al\bar\be}$ to be nonzero at a time.
The direct comparison of the averaged probability \eqref{<Peebar>} with the experimental limit \eqref{Pee_exp} allows us to determine an upper bound on the magnitude of the corresponding coefficient $\gt^{\la\si}_{\al\bar\be}$.
Implementing this procedure,
for the 12 different spacetime indices and six flavor combinations,
72 independent limits are obtained,
where the mixing angles are taken from current global fits \cite{Gonzalez-Garcia:2014,GlobalFits}.
The new upper limits are presented in Table \ref{Table:limits},
together with the existing limits for comparison.

\begin{center}
\begin{table*}[ht] 
\begin{tabular}{cccc|cccc|cccc}
\hline\hline
coefficient & & new limit & previous limit &
coefficient & & new limit & previous limit &
coefficient & & new limit & previous limit \\
\hline
$|  \gt^{XT}_{e\bar e}  |$  & $<$ & $   2.3 \times 10^{-27} $ & $   7.6 \times 10^{-22} $ & $|  \gt^{XT}_{e\bar\mu} |$  & $<$ & $   2.0 \times 10^{-27} $ & $   7.6 \times 10^{-22} $ & $|  \gt^{XT}_{e\bar\tau}    |$  & $<$ & $   3.4 \times 10^{-27} $ & $   8.2 \times 10^{-22} $ \\
$|  \gt^{XX}_{e\bar e}  |$  & $<$ & $   4.7 \times 10^{-27} $ & $   2.0 \times 10^{-21} $ & $|  \gt^{XX}_{e\bar\mu} |$  & $<$ & $   4.0 \times 10^{-27} $ & $   2.0 \times 10^{-21} $ & $|  \gt^{XX}_{e\bar\tau}    |$  & $<$ & $   6.9 \times 10^{-27} $ & $   2.0 \times 10^{-21} $ \\
$|  \gt^{XY}_{e\bar e}  |$  & $<$ & $   2.9 \times 10^{-27} $ & $   2.0 \times 10^{-21} $ & $|  \gt^{XY}_{e\bar\mu} |$  & $<$ & $   2.5 \times 10^{-27} $ & $   2.0 \times 10^{-21} $ & $|  \gt^{XY}_{e\bar\tau}    |$  & $<$ & $   4.3 \times 10^{-27} $ & $   2.1 \times 10^{-21} $ \\
$|  \gt^{XZ}_{e\bar e}  |$  & $<$ & $   6.8 \times 10^{-27} $ & $   1.2 \times 10^{-21} $ & $|  \gt^{XZ}_{e\bar\mu} |$  & $<$ & $   5.7 \times 10^{-27} $ & $   1.2 \times 10^{-21} $ & $|  \gt^{XZ}_{e\bar\tau}    |$  & $<$ & $   1.0 \times 10^{-26} $ & $   1.2 \times 10^{-21} $ \\
$|  \gt^{YT}_{e\bar e}  |$  & $<$ & $   2.2 \times 10^{-27} $ & $   7.6 \times 10^{-22} $ & $|  \gt^{YT}_{e\bar\mu} |$  & $<$ & $   1.8 \times 10^{-27} $ & $   7.6 \times 10^{-22} $ & $|  \gt^{YT}_{e\bar\tau}    |$  & $<$ & $   3.2 \times 10^{-27} $ & $   7.6 \times 10^{-22} $ \\
$|  \gt^{YX}_{e\bar e}  |$  & $<$ & $   2.6 \times 10^{-27} $ & $   2.0 \times 10^{-21} $ & $|  \gt^{YX}_{e\bar\mu} |$  & $<$ & $   2.2 \times 10^{-27} $ & $   2.0 \times 10^{-21} $ & $|  \gt^{YX}_{e\bar\tau}    |$  & $<$ & $   3.9 \times 10^{-27} $ & $   2.0 \times 10^{-21} $ \\
$|  \gt^{YY}_{e\bar e}  |$  & $<$ & $   4.2 \times 10^{-27} $ & $   2.0 \times 10^{-21} $ & $|  \gt^{YY}_{e\bar\mu} |$  & $<$ & $   3.5 \times 10^{-27} $ & $   2.1 \times 10^{-21} $ & $|  \gt^{YY}_{e\bar\tau}    |$  & $<$ & $   6.2 \times 10^{-27} $ & $   2.1 \times 10^{-21} $ \\
$|  \gt^{YZ}_{e\bar e}  |$  & $<$ & $   9.7 \times 10^{-27} $ & $   1.2 \times 10^{-21} $ & $|  \gt^{YZ}_{e\bar\mu} |$  & $<$ & $   8.1 \times 10^{-27} $ & $   1.2 \times 10^{-21} $ & $|  \gt^{YZ}_{e\bar\tau}    |$  & $<$ & $   1.4 \times 10^{-26} $ & $   1.2 \times 10^{-21} $ \\
$|  \gt^{ZT}_{e\bar e}  |$  & $<$ & $   1.7 \times 10^{-27} $ & $   9.7 \times 10^{-18} $ & $|  \gt^{ZT}_{e\bar\mu} |$  & $<$ & $   1.5 \times 10^{-27} $ & $   2.7 \times 10^{-17} $ & $|  \gt^{ZT}_{e\bar\tau}    |$  & $<$ & $   2.5 \times 10^{-27} $ & $   2.7 \times 10^{-17} $ \\
$|  \gt^{ZX}_{e\bar e}  |$  & $<$ & $   2.4 \times 10^{-27} $ & $   1.0 \times 10^{-21} $ & $|  \gt^{ZX}_{e\bar\mu} |$  & $<$ & $   2.0 \times 10^{-27} $ & $   1.0 \times 10^{-21} $ & $|  \gt^{ZX}_{e\bar\tau}    |$  & $<$ & $   3.5 \times 10^{-27} $ & $   1.0 \times 10^{-21} $ \\
$|  \gt^{ZY}_{e\bar e}  |$  & $<$ & $   2.7 \times 10^{-27} $ & $   1.0 \times 10^{-21} $ & $|  \gt^{ZY}_{e\bar\mu} |$  & $<$ & $   2.3 \times 10^{-27} $ & $   1.0 \times 10^{-21} $ & $|  \gt^{ZY}_{e\bar\tau}    |$  & $<$ & $   4.0 \times 10^{-27} $ & $   1.0 \times 10^{-21} $ \\
$|  \gt^{ZZ}_{e\bar e}  |$  & $<$ & $   6.3 \times 10^{-27} $ & $   3.3 \times 10^{-17} $ & $|  \gt^{ZZ}_{e\bar\mu} |$  & $<$ & $   5.3 \times 10^{-27} $ & $   9.3 \times 10^{-17} $ & $|  \gt^{ZZ}_{e\bar\tau}    |$  & $<$ & $   9.2 \times 10^{-27} $ & $   9.3 \times 10^{-17} $ \\
                                                                                            \hline
$|  \gt^{XT}_{\mu\bar\mu}   |$  & $<$ & $   6.6 \times 10^{-27} $ & $   8.9 \times 10^{-24} $ & $|  \gt^{XT}_{\mu\bar\tau}  |$  & $<$ & $   5.8 \times 10^{-27} $ & $   8.6 \times 10^{-24} $ & $|  \gt^{XT}_{\tau\bar\tau} |$  & $<$ & $   2.0 \times 10^{-26} $ & $   1.8 \times 10^{-22} $ \\
$|  \gt^{XX}_{\mu\bar\mu}   |$  & $<$ & $   1.3 \times 10^{-26} $ & $   2.3 \times 10^{-23} $ & $|  \gt^{XX}_{\mu\bar\tau}  |$  & $<$ & $   1.2 \times 10^{-26} $ & $   2.3 \times 10^{-23} $ & $|  \gt^{XX}_{\tau\bar\tau} |$  & $<$ & $   4.1 \times 10^{-26} $ & $   4.8 \times 10^{-22} $ \\
$|  \gt^{XY}_{\mu\bar\mu}   |$  & $<$ & $   8.3 \times 10^{-27} $ & $   2.3 \times 10^{-23} $ & $|  \gt^{XY}_{\mu\bar\tau}  |$  & $<$ & $   7.3 \times 10^{-27} $ & $   2.3 \times 10^{-23} $ & $|  \gt^{XY}_{\tau\bar\tau} |$  & $<$ & $   2.6 \times 10^{-26} $ & $   4.8 \times 10^{-22} $ \\
$|  \gt^{XZ}_{\mu\bar\mu}   |$  & $<$ & $   1.9 \times 10^{-26} $ & $   1.3 \times 10^{-23} $ & $|  \gt^{XZ}_{\mu\bar\tau}  |$  & $<$ & $   1.7 \times 10^{-26} $ & $   1.3 \times 10^{-23} $ & $|  \gt^{XZ}_{\tau\bar\tau} |$  & $<$ & $   5.9 \times 10^{-26} $ & $   2.9 \times 10^{-22} $ \\
$|  \gt^{YT}_{\mu\bar\mu}   |$  & $<$ & $   6.2 \times 10^{-27} $ & $   8.6 \times 10^{-24} $ & $|  \gt^{YT}_{\mu\bar\tau}  |$  & $<$ & $   5.4 \times 10^{-27} $ & $   8.4 \times 10^{-24} $ & $|  \gt^{YT}_{\tau\bar\tau} |$  & $<$ & $   1.9 \times 10^{-26} $ & $   1.8 \times 10^{-22} $ \\
$|  \gt^{YX}_{\mu\bar\mu}   |$  & $<$ & $   7.5 \times 10^{-27} $ & $   2.2 \times 10^{-23} $ & $|  \gt^{YX}_{\mu\bar\tau}  |$  & $<$ & $   6.5 \times 10^{-27} $ & $   2.2 \times 10^{-23} $ & $|  \gt^{YX}_{\tau\bar\tau} |$  & $<$ & $   2.3 \times 10^{-26} $ & $   4.8 \times 10^{-22} $ \\
$|  \gt^{YY}_{\mu\bar\mu}   |$  & $<$ & $   1.2 \times 10^{-26} $ & $   2.2 \times 10^{-23} $ & $|  \gt^{YY}_{\mu\bar\tau}  |$  & $<$ & $   1.0 \times 10^{-26} $ & $   2.2 \times 10^{-23} $ & $|  \gt^{YY}_{\tau\bar\tau} |$  & $<$ & $   3.7 \times 10^{-26} $ & $   4.8 \times 10^{-22} $ \\
$|  \gt^{YZ}_{\mu\bar\mu}   |$  & $<$ & $   2.7 \times 10^{-26} $ & $   1.3 \times 10^{-23} $ & $|  \gt^{YZ}_{\mu\bar\tau}  |$  & $<$ & $   2.4 \times 10^{-26} $ & $   1.4 \times 10^{-23} $ & $|  \gt^{YZ}_{\tau\bar\tau} |$  & $<$ & $   8.4 \times 10^{-26} $ & $   2.9 \times 10^{-22} $ \\
$|  \gt^{ZT}_{\mu\bar\mu}   |$  & $<$ & $   4.9 \times 10^{-27} $ & $   2.3 \times 10^{-16} $ & $|  \gt^{ZT}_{\mu\bar\tau}  |$  & $<$ & $   4.3 \times 10^{-27} $ & $   4.4 \times 10^{-16} $ & $|  \gt^{ZT}_{\tau\bar\tau} |$  & $<$ & $   1.5 \times 10^{-26} $ & $   2.3 \times 10^{-16} $ \\
$|  \gt^{ZX}_{\mu\bar\mu}   |$  & $<$ & $   6.8 \times 10^{-27} $ & $   1.2 \times 10^{-23} $ & $|  \gt^{ZX}_{\mu\bar\tau}  |$  & $<$ & $   5.9 \times 10^{-27} $ & $   1.1 \times 10^{-23} $ & $|  \gt^{ZX}_{\tau\bar\tau} |$  & $<$ & $   2.1 \times 10^{-26} $ & $   2.4 \times 10^{-22} $ \\
$|  \gt^{ZY}_{\mu\bar\mu}   |$  & $<$ & $   7.7 \times 10^{-27} $ & $   1.1 \times 10^{-23} $ & $|  \gt^{ZY}_{\mu\bar\tau}  |$  & $<$ & $   6.7 \times 10^{-27} $ & $   1.2 \times 10^{-23} $ & $|  \gt^{ZY}_{\tau\bar\tau} |$  & $<$ & $   2.4 \times 10^{-26} $ & $   2.4 \times 10^{-22} $ \\
$|  \gt^{ZZ}_{\mu\bar\mu}   |$  & $<$ & $   1.8 \times 10^{-26} $ & $   8.1 \times 10^{-16} $ & $|  \gt^{ZZ}_{\mu\bar\tau}  |$  & $<$ & $   1.5 \times 10^{-26} $ & $   1.5 \times 10^{-15} $ & $|  \gt^{ZZ}_{\tau\bar\tau} |$  & $<$ & $   5.4 \times 10^{-26} $ & $   8.1 \times 10^{-16} $ \\
\hline\hline
\end{tabular} 
\caption{ New upper limits on the magnitude of 72 independent coefficients for CPT violation within the SME. 
These results can also be interpreted as limits on the real and imaginary parts of the coefficients.
The flavor indices for each coefficient take six different combinations,
and spacetime indices take 12 combinations.
Existing limits on these dimensionless coefficients are also shown for comparison \cite{Diaz:2013b,Rebel:2013}. }
\label{Table:limits}
\end{table*}
\end{center}

The results in Table \ref{Table:limits} show a remarkable improvement on the limits for coefficients for CPT violation.
The most notorious improvements are on the coefficients with spacetime components $\la\si=ZT,ZZ$,
whose existing bounds were obtained by searching for spectral distortions in the Double Chooz data \cite{Diaz:2013b}.
In this reactor experiment, 
antineutrinos only travel about 1 km,
which explains the great improvement by factors $10^9$--$10^{11}$ when using solar neutrinos.
All the other coefficients produce sidereal variations,
which were studied with data from MINOS \cite{Rebel:2013}.
For these coefficients,
the 735-km propagation distance enhances the sensitivity of the experiment by almost 3 orders of magnitude with respect to Double Chooz.
Moreover,
neutrinos in MINOS have a thousand times more energy than reactor antineutrinos,
which also increases the sensitivity to $\gt^{\la\si}_{\al\bar\be}$.
Compared to MINOS,
solar neutrinos have only one thousandth of the energy,
but again the Sun-Earth distance compensates and the sensitivity of solar neutrinos to these coefficients is much higher.
As shown in Table \ref{Table:limits},
most of the limits on these coefficients are improved by at least 3 orders of magnitude.
Since the relevant oscillation channel in MINOS involves $\nu_\mu$ and $\nu_\tau$,
its sensitivity was low for coefficients with electron-flavor components.
For this reason,
solar neutrinos are up to a factor $10^5$ more sensitive to these coefficients.

\section{Summary}
\label{Sec.Summary}

Existing searches for deviations from exact Lorentz invariance in the neutrino sector have been performed by different collaborations using a variety of accelerator, atmospheric, and reactor experiments \cite{LV_LSND, LV_MINOS_ND1, LV_IceCube, LV_MINOS_FD, LV_MiniBooNE:2012, LV_DoubleChooz, LV_MINOS_ND2, LV_MiniBooNE:2013,LV_SuperK}.
Many of the coefficients controlling the unconventional effects in these experiments have been constrained. 
For instance, 
the sensitivity reach on the dimensionless coefficient for CPT-even Lorentz violation $(c_L)^{\la\si}_{\ab}$ span the $10^{-16}$--$10^{-27}$ range,
where the high sensitivity end is due to the high energy and long propagation distance of atmospheric neutrinos studied by IceCube \cite{LV_IceCube} and Super-Kamiokande \cite{LV_SuperK}.
These limits could be improved in the future by systematic studies of the observed flavor ratios of astrophysical neutrinos \cite{Arguelles:2015}.
Some components of these coefficients that leave flavor mixing unaffected can also be studied with astrophysical neutrinos \cite{Diaz:2014a}.

The phenomenon of neutrino-antineutrino mixing has been explored using reactors and accelerators \cite{Diaz:2013b,Rebel:2013};
nonetheless,
the sensitivity of these terrestrial studies is mostly limited by the neutrino propagation distance.
In the present work,
we have taken advantage of the remarkable propagation distance traveled by neutrinos from the Sun.
Even though the sensitivity to the effects of the coefficients for CPT violation $\gt^{\la\si}_{\al\bar\be}$ is partially suppressed by the low energy ($\sim$MeV) of $^8$B neutrinos,
the vast Earth-Sun separation is millions of times larger than the standard baseline in laboratory experiments.
This advantage of solar neutrinos allowed us to improve the limits on coefficients for CPT violation $\gt^{\la\si}_{\al\bar\be}$ 
by factors ranging from about a thousand to $10^{11}$.
Our results allow pushing the sensitivity reach on each of the 72 coefficients $\gt^{\la\si}_{\al\bar\be}$ to the $10^{-26}$--$10^{-27}$ range.
Unfortunately,
the coefficients for CPT-even Lorentz violation $\Ht^{\la}_{\al\bar\be}$ are unobservable in the oscillation channel studied in the present work and the limits on these coefficients remain in the $10^{-18}$--$10^{-22}$ GeV range.

\section*{Acknowledgments}
The work of J.S.D. was supported in part by the German Research Foundation (DFG) under Grant No. KL 1103/4-1.


%
\end{document}